\documentclass[12pt,twocolumn]{iopart}
\usepackage{graphicx,gensymb}

\usepackage{bm}

\begin{document}

\title{Structural properties of crumpled cream layers}

\author{M A F Gomes$^{1}$\footnote{Corresponding author e-mail: mafg@ufpe.br; fax: ++55 81
3271 0359;  phone: ++55 81 2126 7614}, C C Donato$^1$, S L
Campello$^2$, R E de Souza$^{1,2}$, and R Cassia-Moura$^{3,4}$}

\address{$^1$Departamento de F\'{\i}sica, Universidade Federal de
Pernambuco, 50670-901, Recife - PE - Brazil}

\address{$^2$Programa de P\'os-Gradua\c{c}\~ao em Ci\^encia de
Materiais, Universidade Federal de Pernambuco, 50670-901, Recife,
PE, Brazil}

\address{$^3$Instituto de Ci\^encias Biol\'ogicas,
DCF-Biof\'{\i}sica, Universidade de Pernambuco, Caixa Postal 7817,
50670-000, Recife, PE, Brazil}

\address{$^4$International Centre
for Theoretical Physics, Trieste 34100, Italy}

\begin{abstract}

The cream layer is a complex heterogeneous material of biological
origin which forms spontaneously at the air-milk interface. Here,
it is studied the crumpling of a single cream layer packing under
its own weight at room temperature in three-dimensional space. The
structure obtained in these circumstances has low volume fraction
and anomalous fractal dimensions. Direct means and noninvasive NMR
imaging technique are used to investigate the internal and
external structure of these systems.

\end{abstract}

\pacs{68.18.Fg, 89.75.Da, 87.61.-c, 05.40.-a}

\maketitle

\section{\label{sec1} Introduction}

\indent The crumpling of a surface, a sheet polymer, or a
self-avoiding elastic model sheet is a subject of great interest
in both theoretical~\cite{1a,1b}, and applied physics~\cite{2,3},
as well as in materials science~\cite{4}, due to the peculiar
properties of the crumpled structures. It is known that both a
paper sheet and a thin metal foil of area $A$ and mass $M$ when
submitted to a haphazard ill-defined compaction by manual means in
order to generate an approximately globular object, present a
non-thermal transition to a crumpled fractal state of packing
characterized by the formation of a complex pattern of folds. The
crumpled structures obtained with these materials obey the
nontrivial scaling

\begin{equation}
A \sim M \sim \phi^{D_M}, \label{eq1}
\end{equation}

\noindent where $\phi$ is the corresponding average globular
diameter of the structure after crumpling, and $D_M$ is the mass
fractal dimension~\cite{2,5,6}. This fractal dimension assumes in
these cases values in the interval $2.2 < D_M < 2.7$ irrespective
of the material, and of the thickness of the foil in the interval
of 20-200 $\mu$m studied~\cite{2,5,6}.

\indent Anyone knows from childhood that from ordinary milk we can
obtain the cream layer, that tenuous whitish membranous layer
formed on the upper part of the cream right at the air-milk
interface. The cream layer obtained from milk is a thin sheet that
contains micron- and submicron-sized fat globules, proteins,
phospholipids, and water, among other components, and is formed
very rapidly, within a few minutes, in cold milk. The cream layer
is formed at the free surface of milk as a planar quasi
two-dimensional continuous internal fat network~\cite{7,8} as the
result of a complex process of coalescence of the fat globules.
Milk is in fact one of the most complex foods, with more than
$100,000$ different molecular components~\cite{7}. Besides its
economic importance in dairy industry, milk is the most important
source of nourishment and immunological protection for young
mammals, and for humans it has been a food source since
prehistoric times.

\indent Here we investigate the crumpling associated with the
three-dimensional packing of a single cream layer. Our extensive
analysis indicates that the cream layer collapses under its own
weight at room temperature into a three-dimensional fractal
structure of low volume fraction. The structure of this paper is
the following: in the Section~\ref{sec2} we describe the
experimental details, in Section~\ref{sec3} we present our
results, and in Section~\ref{sec4} we have a summary of our
conclusions.

\section{\label{sec2} Experimental Details}

\indent In the first place, it is interesting to note that a sheet
of paper or a thin metal foil of area $A$ is unable to
spontaneously crumple fractally under the action of its own
weight, for all practical values of $A$. On the other hand, from
the point of view of a system at thermal equilibrium, it has been
suggested that sheet polymers can exhibit a crumpling transition
satisfying the fractal scaling (\ref{eq1}) above a particular
critical temperature~\cite{9}. This last transition, however, had
not yet been experimentally observed. In the present study, using
direct means and noninvasive imaging techniques, we study the
crumpled state observed when a cream layer is slowly deposited
onto a glass support, and packs fractally under the action of its
own weight, forming a fresh crumpled cream layer (CCL). A fresh
CCL in the context of the present work has high water content,
i.e. it is made and analyzed within a few hours after its
transference from the milk free surface to the surface where it is
deposited and studied. The total time involved in the stages of
transference and deposition of each cream layer is typically
$30-40$~s. This period includes a stage of drainage in which all
excess liquid is lost. All the measurements reported here were
made at room temperature of $(24 \pm 1)$~$^\circ$C.

\indent We have made 90 circular planar layers, and from these we
selected 72 specimens to study. The criteria used in this
selection were uniformity of the cream layer and mechanical
resistance to manipulation. By heating UHT (ultra heat treated)
cow milk to boiling point, under agitation, cream layers with
2.0-28.5~cm of diameter were spontaneously made. Immediately after
it began to boil, the milk was distributed in circular containers
of different diameters without foam formation. After approximately
2 minutes at rest, a visible random shrinking of the layer, in
which the edges of the layer come loose from the container,
follows the spontaneous layer formation. As a result, systematic
variability arises from successive identical trials, i.e. layers
of almost equal diameter may have been obtained in different
containers, and one container can generate distinct values of
layer diameter.

\begin{figure*}[!]

\centering {\includegraphics{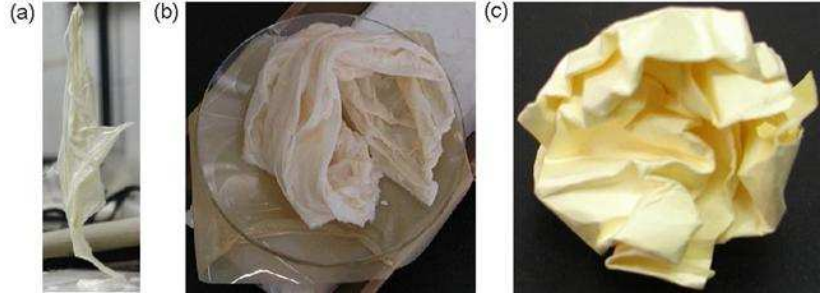}}

\caption{\label{fig1} (a) When the cream layer is slowly deposited
onto a horizontal surface, it progressively crumples into a
three-dimensional structure under the action of its own weight. In
this case (CCL$\#1$), the original cream layer had an area of
638~$\mbox{cm}^2$, and an average thickness of 82~$\mu$m. (b)
Photograph of the final aspect of CCL$\#1$ with a diameter $\phi =
3.1$~cm. The packing process is arrested in a fractal state of low
volume fraction $\eta \simeq 0.33$. (c) A crumpled surface of
paper 125~$\mu$m thick with size and fractal dimension close to
CCL$\#1$.}
\end{figure*}

\indent For illustration figure~\ref{fig1}(a) shows a photograph
taken during the operation of deposition of CCL$\#1$. Each layer
was separated from the milk and softly deposited onto the support
with the aid of a slightly curved wire handle. The packing process
of the cream layer begins when it is rising from the milk surface
with the aid of the wire handle: the layer presents its first fold
and approximately assume the aspect of a vertical half circle with
twice the thickness of the original (horizontal) cream layer. This
semicircular layer is highly unstable: it flows along the
curvature of the wire handle in order to minimize its
gravitational potential energy, and assumes the aspect of a
many-folded strip, as shown in figure~\ref{fig1}(a). While the
layer is slowly deposited, it progressively crumples into a
three-dimensional structure under the action of its own weight. In
this particular case, the original planar cream layer had a
diameter of approximately 28.5~cm, and an average thickness of
82~$\mu$m. This last numerical value is an estimate based on the
mass, density and radius of the planar fresh cream layer.
Figure~\ref{fig1}(b) shows the external appearance of the cream
layer after crumpling (CCL$\#1$). The three-dimensional diameter
of the CCL$\#1$ was 3.1~cm, and the volume fraction $\eta \simeq
0.33$, i.e. approximately $51\%$ of the volume fraction for the
random close packing of spheres in three-dimensions~\cite{10}. For
comparison, the photograph of a crumpled surface of paper of
similar external size and thickness of 125~$\mu$m is shown in
figure~\ref{fig1}(c).

\section{\label{sec3} Results and Discussion}

\begin{figure*}[!]

\centering {\includegraphics{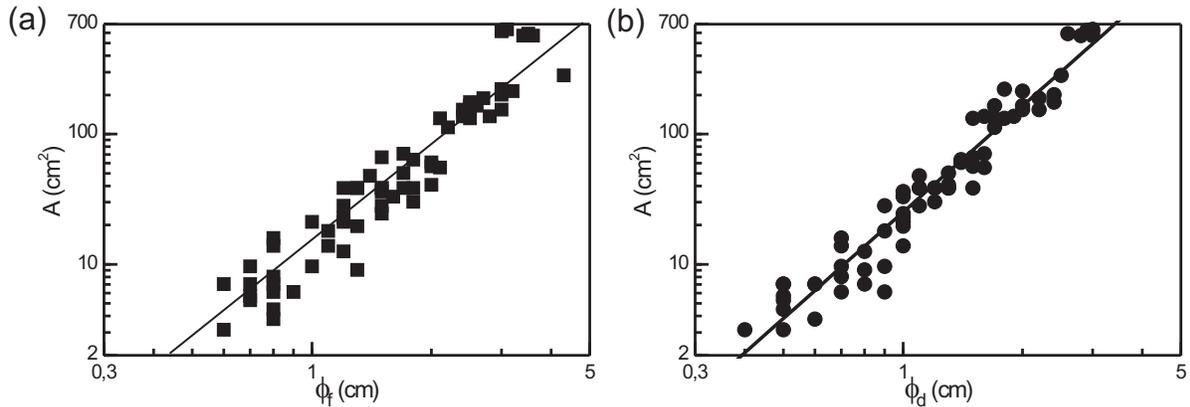}}

\caption{\label{fig2} Area $A$ of the cream layer as a function of
the external three-dimensional diameter $\phi$: (a) For the
ensemble of 72 fresh CCL (i.e. $\phi$ measured immediately after
transference of the cream layer from milk-free surface to the
glass support): the continuous line refers to the best fit $A \sim
\phi_f^{2.45 \pm 0.15}$. (b) For the corresponding ensemble of 72
dry CCL (with lower water content: $\phi$ measured 10 days after
CCL formation): $A \sim \phi_d^{2.65 \pm 0.10}$.}
\end{figure*}

\indent There are various ways to characterize the basic
structural properties of real irregular surfaces~\cite{11,12}.
Here, we choose to investigate the geometric properties of
crumpled cream layers in terms of both the mass fractal dimension
and the box dimension. Firstly, we measured the dependence of the
area $A$ with the average three-dimensional diameter $\phi$
(equation~(\ref{eq1})) for the ensemble of 72 CCL. The
experimental samples of cream layers studied here had an area
(average diameter) in the interval $3 \leq A(\mbox{cm}^2) \leq 638
\; (0.4 \leq \phi(\mbox{cm}) \leq 4.3)$, and a surface thickness
$\zeta  = (82 \pm 9)$~$\mu$m. The $A \times \phi$ measurements
give rise to the plots shown in figures~\ref{fig2}(a) and
\ref{fig2}(b), respectively, for fresh ($f$) samples, as well as
for dry ($d$) rigid samples examined 10 days after CCL formation,
and consequently with low water content. The continuous lines in
these plots refer to the corresponding best fits $A \sim
\phi_f^{2.45 \pm 0.15}$, and $A \sim \phi_d^{2.65 \pm 0.10}$, i.e.
there is a slight increase in the mass fractal dimension (and a
slight reduction in the statistical fluctuations) with the aging
of the system, that is when the system evolves from fresh CCL to
dry CCL. The external measurements of the diameter of the CCLs
suggest that after crumpling these systems condense in the
three-dimensional physical space as an anomalous non-space-filling
structure with mass fractal dimension $D_M < 3$. However, external
measurements of diameter and area are not the sole way to support
the claim of a fractal state for CCLs, so we have used another
tool to test the crumpling properties of these low volume fraction
structures.

\indent Secondly, to obtain a more detailed microscopic picture of
the three-dimensional internal structure of the CCLs we decided to
investigate these systems using NMR imaging~\cite{13}. In this
case, NMR imaging technique is unique to reveal the internal
structure of the CCL samples. Here we evaluate the
three-dimensional mass distribution of CCL samples from
measurements of proton density maps. Imaging experiments were
performed on a (Varian) Unity Inova spectrometer that includes a
2.0~T, 30~cm horizontal room temperature bore magnet. All
experiments were carried out at ($24 \pm 1$)~$^\circ$C. Spin-spin
relaxation time $T_2$ of protons in CCL ranges from a hundred
microseconds to the extreme-narrowing condition where $T_2$ value
reaches hundreds of milliseconds. Since CCL is a highly hydrated
material (liquid water content $> 60\%$), most of the protons
belong to two categories: free water protons, and water protons in
exchange with exchangeable protein protons~\cite{14}. In order to
measure the mass distribution more accurately we have used a
three-dimensional gradient spin echo pulse sequence, where the
short non-selective $\pi/2$~RF pulse allows measurement of spin
echoes before severe $T_2$ attenuation takes place. Given that the
longest spin-lattice relaxation time $T_1$ measured in the control
sample was $\sim 300$~ms, a recycling time $T_R = 1$~s  was
employed in order to minimize corrections due to the $T_1$
relaxation process. Mostly, we have used an imaging matrix of $128
\times 64 \times 64$, slice thickness of 1~mm, and $T_E \geq
1.2$~ms. Each three-dimensional image took 68 minutes. Since
experiments are short in time, no appreciable loss of liquid by
the samples is observed. After all the corrections we can state
that pixel intensities on the images are proportional to free and
almost-free water content, and to protons in small molecules.
Finally, it is possible that NMR images might not represent the
total mass distribution of CCL exactly. If this is the case, our
estimate for the ensemble average of the box fractal dimension for
CCLs could be somewhat underestimated.

\begin{figure*}[!]

\centering {\includegraphics{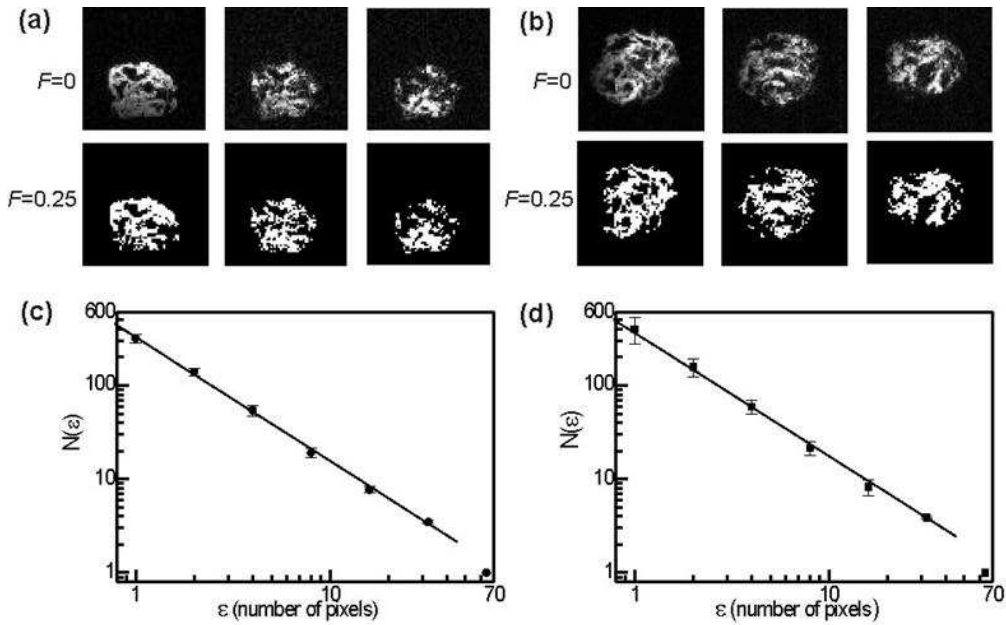}}

\caption{\label{fig3} Typical NMR images of cross sections of
fresh CCL$\#1$ and box-counting plots: (a) Plane $xy$ (parallel to
the gravitational field). (b) Plane $xz$ (orthogonal to the
gravitational field). The images have $64^2$ pixels and refer to
the absence of any filtering on the signal ($F = 0$), and to a
filtering factor of $25\% \; (F = 0.25)$. (c) and (d) show the
corresponding box-counting plot~\cite{15} that gives the number
$N(\epsilon)$ of boxes of size $\epsilon \times \epsilon$ needed
to cover the image (for $F = 0.25$). The plots give the scaling
$N(\epsilon) \sim \epsilon^{-\delta}$ (continuous line), with
$\delta = 1.30 \pm 0.05$. The fractal dimension of the CCL$\#1$ in
three-dimensional physical space is $D = \delta + 1 = 2.30 \pm
0.05$. Fluctuation bars are associated with means on different
(parallel) image planes and experimental tuning parameters.}
\end{figure*}

\indent Thus images of two-dimensional sections of large fresh
CCLs were obtained along three orthogonal planes $xy$, $yz$ (both
containing the direction of the gravitational field), and $xz$
(orthogonal to the gravitational field), and in intervals
separated by a distance of 1~mm. Figures~\ref{fig3}(a) and
\ref{fig3}(b) (\ref{fig3}(c) and \ref{fig3}(d)) show the NMR
images (and the corresponding box-counting analysis~\cite{15} that
gives the number $N(\epsilon)$ of square boxes of size $\epsilon
\times \epsilon$ needed to cover the image) along, respectively,
some intermediary planes $xy$, and $xz$ for CCL$\#1$, for three
different values of the experimental tuning parameters. The images
refer to the situation of absence of any filtering on the signal
($F = 0$), and with a filtering factor of $25\% \; (F = 0.25)$. In
figures~\ref{fig3}(a) and~\ref{fig3}(b) each one of the three
columns refer to a particular set of experimental tuning
parameters. Both box-counting plots are well described by the
scaling $N(\epsilon) \sim \epsilon^{-\delta}$ along $1.5$ decades
in $\epsilon$, as can be seen from the power law best fits
(continuous straight lines in figures~\ref{fig3}(c) and
\ref{fig3}(d)), both with a slope of $\delta = 1.30 \pm 0.05$.
This value of $\delta$ represents the box fractal dimension of
those particular sections~\cite{15}. The corresponding box fractal
dimension of the CCL in three-dimensional physical space~\cite{15}
is $D = \delta + 1 = 2.30 \pm 0.05$. The fluctuation bars in
figures~\ref{fig3}(c) and ~\ref{fig3}(d) are associated with the
means on the different image planes and experimental tuning
parameters. In general, the value of $D$ for any sample is robust
within typical statistical fluctuations of $3\%$ to $5\%$,
irrespective the value of $F$ and the other experimental
parameters.

\begin{figure}[h!]
\centering \resizebox{9cm}{!}{\includegraphics{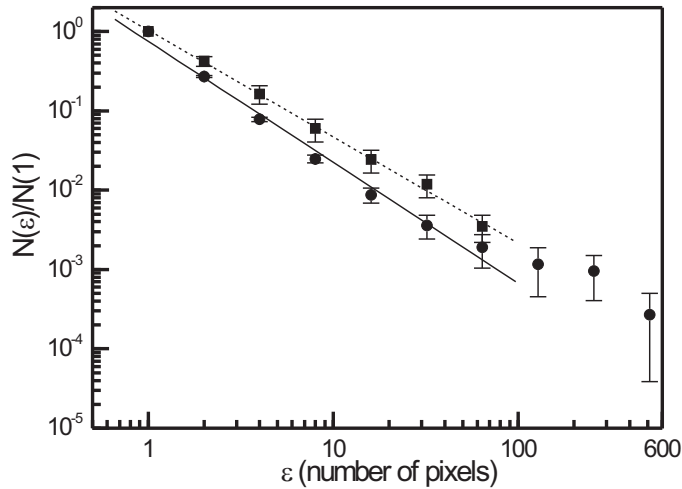}}
\caption{\label{fig4} Box-counting plot for large fresh CCLs, for
$F = 0.25$. The dashed (continuous) line is associated with
CCL$\#1$ (CCL$\#2$). Our overall estimate for the box dimension of
the ensemble of CCL is $D_{ens} = 1 + \delta_{ens} = 2.38 \pm
0.10$.}
\end{figure}

In figure~\ref{fig4} we show the box-counting plot for the entire
ensemble of CCL images studied: 550 experimental data points for
$F = 0.25$, after averaging on image samples, on three orthogonal
image planes ($xy$, $yz$ and $xz$), and on experimental tuning
parameters. The slopes of the straight lines associated with the
power law fits are $\delta_{64} = 1.34 \pm 0.03$, for $439$ images
with $64^2$ pixels (dashed line) from CCL$\#1$, and $\delta_{426}
= 1.53 \pm 0.07$, for $111$ images with $426^2$ pixels (continuous
line) from CCL$\#2$ ($A = 625~\mbox{cm}^2$, $\phi = 3.0$~cm, and
$\eta \simeq 0.36$). CCL$\#1$ and CCL$\#2$ were the two largest
samples in the ensemble studied. The fluctuations bars are due to
the mean on many orthogonal image planes and on experimental
tuning parameters. The weighted ensemble mean of these values
gives the result $\delta_{ens} = 1.38 \pm 0.10$. No noticeable
change in $\delta_{ens}$ was observed within practically the full
interval of variation of $F$. Thus, our overall estimate for the
ensemble average of the box fractal dimension of CCL is $D_{ens} =
\delta_{ens} + 1 = 2.38 \pm 0.10$, which is equal within the
fluctuation bars, to the mass dimension obtained from
figure~\ref{fig2}(a), and from experiments with crumpled sheets of
paper~\cite{5,6}, metal foils~\cite{2}, as well as from computer
simulations~\cite{4}, and a Flory-type approximation expected to
be a value for membranes at thermal equilibrium~\cite{9}.

\section{\label{sec4} Conclusions}

\indent Our results, based on a large number of experiments of
deposition of thin cream layers of cow milk onto a glass support
have shown that, although the cream layers collapse under their
own weight at room temperature, they never collapse into a
three-dimensional compact structure. Rather, the packing process
of these tenuous CCL structures is arrested in an intermediary
crumpled state of low volume fraction. This crumpled state is
surprisingly rigid in the case of dry samples. Two important
experimental aspects of this low-density state are its anomalous
area-size scaling, and its anomalous box dimension. To the best of
our knowledge, it is the first time that noninvasive NMR imaging
is used to obtain information on the interior of crumpled surfaces
(figures~\ref{fig3} and~\ref{fig4}). Moreover, the measured
fractal dimension of a CCL, a system of animal origin, is equal to
that observed for other crumpled surfaces made of completely
different materials and obtained by completely different
means~\cite{2,5,6}. These findings suggest that a universal
dynamics may be responsible for all these crumpling processes. In
particular, the results reported here seem to confirm that the
crumpling dynamics is heavily dependent on a few attributes of the
system, as exemplified by the two-dimensional topology of the
surfaces~\cite{2}. In conformity with a recent work~\cite{16}, and
perhaps most importantly, our results indicate that to a large
extent, the fractal dimension of the crumpled surfaces does not
depend on the magnitude of attractive interactions transverse to
the layer, which are expected to exist in the case of CCL but are
absent for metal foils and sheets of paper. The robustness of the
numerical value of the mass fractal dimension observed in the
macroscopic crumpling experiments reported here for CCL, and in
other works for different materials and conditions~\cite{2,5,16},
may be an indicative that a similar type of anomalous packing can
be found when the size of the surfaces is reduced to
$\mbox{micron}^2$ area scales.

\section*{Acknowledgments}

This work was supported in part by Conselho Nacional de
Desenvolvimento Cient\'{\i}fico e Tecnol\'ogico (CNPq), Programa
de N\'ucleos de Excel\^encia, and N\'ucleo de Materiais
Avan\c{c}ados (Brazilian government agencies.). C.C.D.
acknowledges a postdoctoral fellowship from CNPq. We are grateful
to E. N. Azevedo for his technical assistance with part of the NMR
images. M.A.F.G. expresses his thanks to G. L. Vasconcelos, and I.
R. Tsang for their fruitful discussions.

\end{document}